\documentclass[conference, a4paper]{IEEEtran}
\IEEEoverridecommandlockouts

\usepackage[top=19mm, bottom=43mm, left=13mm, right=13mm]{geometry}
\usepackage{cite}
\usepackage{amsmath,amssymb,amsfonts,mathtools}
\usepackage{algorithm}
\usepackage{algorithmic}
\usepackage{graphicx}
\usepackage{textcomp}
\usepackage{xcolor}
\usepackage{booktabs}
\usepackage{tabularx}
\usepackage{siunitx}
\usepackage{multirow}
\usepackage{bm}
\usepackage{epstopdf}

\makeatletter
\def\ps@IEEEtitlepagestyle{
  \def\@oddhead{}
  \def\@evenhead{}
  \def\@oddfoot{\footnotesize\textsf{979-8-3195-4565-7/26/\$31.00~\copyright~2026 IEEE}\hfill}
  \def\@evenfoot{}
}
\begin{document}

\title{Deep Learning-Based Multi-User Communication Design for Dense IoT Networks: Interference-Aware Finite-Blocklength Communication and Preliminary MIMO Extensions}

\author{
\IEEEauthorblockN{Arkadeep Sinha\textsuperscript{*}, Shubham Paul\textsuperscript{*}, R.~Manivasakan\textsuperscript{*}}
\IEEEauthorblockA{Email: paulshubham96@outlook.com, arkadeepsinha98@gmail.com\\
\textsuperscript{*}Indian Institute of Technology Madras}
}

\maketitle
\begin{abstract}
Dense IoT networks require reliable communication despite limited spectrum and substantial multi-user interference while maintaining manageable receiver complexity. This work introduces a deep-learning-based end-to-end multi-user communication design for interference-limited finite-blocklength IoT scenarios, focusing on short and medium blocklengths.  

We extend a prior 2-user SiameseNet transceiver framework to accommodate 2, 4, and 8 users, leveraging learned redundancy for interference suppression and noise robustness. Compared to conventional non-orthogonal access baselines, our method demonstrates strong Block Error Rate (BLER) performance across various scenarios without resorting to joint detection; the per-user decoder scales roughly linearly with the number of users.

Further, we examine the robustness under interference mismatch and unequal interference strengths, critical for practical  deployments with heterogeneous devices. The Latent-space analysis reveals that the learned codeword distance increases as the effective per-user rate decreases, corroborating with the observed BLER improvements. In addition, we also present preliminary results for a 2X2 MIMO setup under fixed-channel CSIT and CSIR, indicating potential for extending the framework to IoT gateways with multiple antennas. 

\end{abstract}
\IEEEpubidadjcol

\section{Introduction}
\label{sec: Introduction}

In dense Internet of Things (IoT) networks, a large number of connected devices transmit finite-blocklength packets while managing a stringent spectrum, latency, and reliability constraints. These packets, often short, may also differ in blocklength depending on payload size and access conditions. Consequently, multi-user interference emerges as a significant bottleneck, which requires communication strategies that enhance noise robustness, manage interference, and reduce decoding complexity in finite blocklengths.

Conventional methods limit the interference by orthogonalising resources (TDMA, FDMA, CDMA) and employing robust Forward Error Correction (FEC) like Polar ~\cite{Polar_Code}, LDPC ~\cite{LDPC_Gallager} and Turbo codes to meet error targets.

While these approaches are effective, these strategies can reduce spectral efficiency .
Alternatives such as Non-Orthogonal Multiple Access (NOMA~\cite{NOMA_OG}) and Rate-Splitting Multiple Access (RSMA) ~\cite{Mao_2018, Outage_Downlink_RSMA, Primer_Rate_Splitting_Multiple_Access} enhance spectral efficiency but often introduce complexities in joint detection~\cite{Int_cancel_NOMA, NOMA_Decoding, Low_Complexity , Coop_RA_NOMA, Modulation_Coding_NOMA, Zero_SIC}, presenting scalability challenges, especially within low-complexity, scalable IoT environments.


DL allows End-to-End (E2E) physical-layer optimisation, integrating modulation, coding, and decoding, along with robustness techniques such as variational inference and backpropagation through non-differentiable channels~\cite{Variational_Inference, DL_Air}. For NOMA, DL has supported joint detection and MIMO-SIC optimisation~\cite{Joint_Detection_NOMA, MIMO_NOMA_SIC_2020}. However, prior work has not focused on E2E optimisation in high-interference regimes.

 For Z-interference channels, a deep autoencoder (DAE-ZIC) was benchmarked in~\cite{Autoencoder_Z_intf}. TwinNet and SiameseNet~\cite{Paul_Intf_lim} leveraged interference for performance gains but were limited to two users. 
Our research work targets scenarios relevant to dense IoT connectivity, where multiple devices simultaneously compete for reliable transmission amid mutual interference. We assess Block Error Rate (BLER) performance as well as robustness against incorrect interference assumptions and varying interference strengths, analysing learned codeword geometry for insights into performance behaviour.

The main contributions of this paper are as follows:

\begin{itemize}
\item We generalise the SiameseNet transceiver model in ~\cite{Paul_Intf_lim}, from 2-user scenario to 4-user and 8-user interference-limited settings
and provide a systematic BLER evaluation across these regimes.
\item We investigate the robustness against practical mismatches, including mismatched interference strength and unequal interference strength assumptions.
 \item We analyse learned latent codeword geometry through minimum-distance and correlation measures to elucidate observed BLER trends.
 \item We compared our proposed approach and results to both orthogonal and non-orthogonal baselines, discussing associated decoding complexity trade-offs.
 \item We also provide preliminary 2×2 MIMO results under fixed-channel assumptions as a foundational step beyond single-input single-output (SISO) contexts.
 
\end{itemize}



The paper is structured as follows: Section~\ref{sec: System Model} presents the System Model, Section ~\ref{sec: Introduction_to_Siamese_Architecture} details the Siamese architecture and Decoding Complexity, Section ~\ref{sec: BLER Performance} reports BLER Performance, Section ~\ref{sec: Integration with MIMO} provides preliminary Results with MIMO, and Section~\ref{sec: Latent Space Analysis} Analyses the Latent Codewords and Conclusions are drawn in Section~\ref{sec: conclusions}. 

\section{System Model}\label{sec: System Model}
\begin{figure}[t]
\centering
\includegraphics[width=0.9\linewidth]{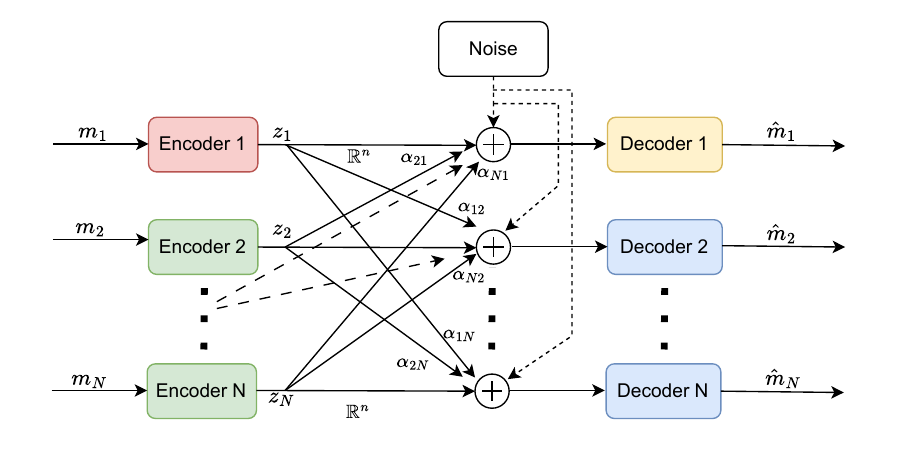}
\caption{Multi-user interference system model for $N$ users.}
\label{fig: System model 2}
\end{figure}

The Siamese network of~\cite{Paul_Intf_lim} considered two users with symmetric interference. Here we extend it for $N\in\{4,8\}$ users. The interference model is shown in Fig.~\ref{fig: System model 2}. The dashed lines indicate interferers not explicitly drawn. The interference from user $j$ to user $i$ has magnitude $\alpha_{ij}$, with $\alpha_{ij}=\alpha_{ji}$.

As in~\cite{Paul_Intf_lim}, user~$i$ maps a $k$-bit message $M_i\in\{0,1\}^k$ to a real $n$-length codeword $z_i$ via
\begin{gather}
E_i \equiv F_{\theta_i}: M_i \rightarrow z_i , \quad i \in \{1,2,\dots,N\},
\label{eq: Encoder_eqn}\\
\mathbb{E}\!\left[\lVert z_i \rVert_{2}^{2}\right] = n,
\label{eq: Power Cons}
\end{gather}
where $\theta_i$ are the encoder parameters. The received vector at $Rx_i$ is
\begin{equation}
y_i = z_i + \sum_{\substack{j=1\\ j\neq i}}^{N} \alpha_{ij}\, z_j + w_i,\quad w_i \sim \mathcal{N}(0, \sigma^2).
\label{eq: rx_eqn}
\end{equation}
where $w_i$ denotes white noise samples.
Decoder $D_i$ estimates $\hat{M}_i$ from $y_i$:
\begin{gather}
D_i \equiv G_{\phi_i}:  y_i \rightarrow \hat{M}_i , \quad i \in \{1,2,\dots,N\}, \label{eq: Dec_eq}\\
P_{e} = \frac{1}{N}\sum_{j=1}^{N} \Pr\!\left(\hat{M}_j\neq M_j\right),\label{error_prob}
\end{gather}
where $\phi_i$ are decoder parameters.

\section{Siamese Architecture and Decoding Complexity}
\label{sec: Introduction_to_Siamese_Architecture}
\subsection{Model Structure and Design}
\begin{figure}[htbp]
\centerline{\includegraphics[width=0.9\linewidth]{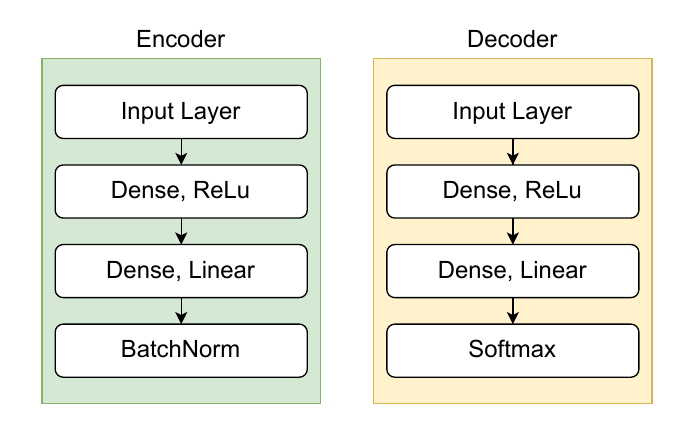}}
\caption{Encoder and Decoder Structure}
\label{fig: Encoder and Decoder Structure}
\end{figure}

\begin{figure}
  \centering
  \includegraphics[width=0.9\columnwidth]{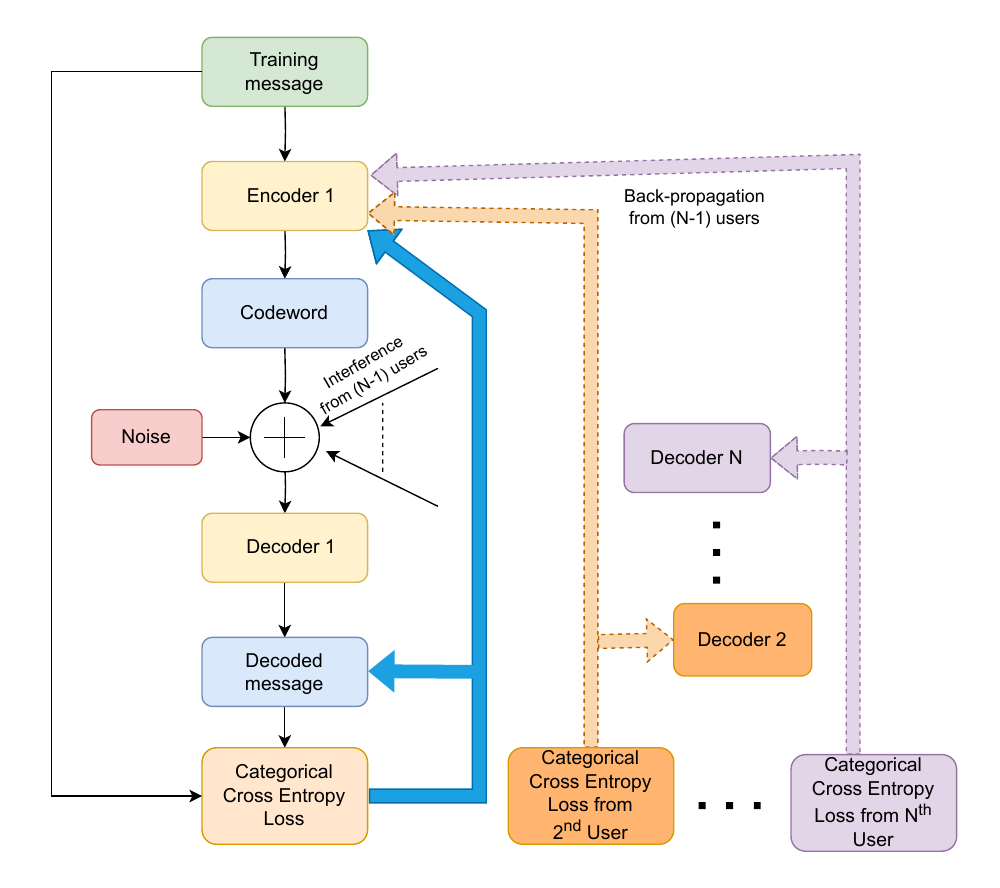} 
  \caption{The Siamese Network Architecture and Training (extended to N users).}
  \label{fig:Siamese_arch}
\end{figure}

Following the algorithm given in~\cite{Paul_Intf_lim}, interfering users learn from errors observed at both decoders. We modify the algorithm given in~\cite{Paul_Intf_lim} to allow interference from all users. 

The autoencoder comprises an encoder, a non-trainable channel layer (latent perturbations), and a decoder. As shown in Fig.~\ref{fig: Encoder and Decoder Structure}, the encoder processes a one-hot input through a ReLU dense layer, a linear layer, batch normalisation, and an additive Gaussian layer (AWGN surrogate). The decoder uses a ReLU hidden layer, a linear layer, and a softmax output. Interference is added at the channel input.

For $N$ users and blocklength $L$, the hidden and encoder output (BatchNorm) sizes are $NL$. The decoder has one ReLU hidden layer of size $NL$, followed by a linear layer and softmax. Interference from other users is injected at the current user’s channel input. We encourage the reader to refer~\cite{Paul_Intf_lim} for further details about the model.

Fig. 3 shows the Siamese network architecture extended to
N users. A small change in $\theta_i$ affects all costs $\{C_j\}_{j=1}^N$, enabling joint encoder learning, as demonstrated in Algorithm~\ref{alg:Siamese-network}. In contrast, small perturbations in $\phi_i$ only affect $C_i$, so decoders are learned independently. This is indicated by backpropagation arrows: losses from User~2 feed only Decoder~2 and Encoder~1, not Decoder~1 or Decoder~$N$.

\begin{algorithm}[t]
\caption{Training the Siamese network}
\label{alg:Siamese-network}
\begin{algorithmic}[1]
\WHILE{epoch $<$ maxepochs}
\STATE $\frac{E_b}{N_0} \gets \mathrm{Uniform}[1,12]\ \mathrm{dB}$
\STATE $w_i \gets \mathrm{AWGN}\!\left(\sigma\!\left(\tfrac{E_b}{N_0}\right)\right)$ for $i=1,\dots,N$
\STATE $z_i \gets E_{\theta_i}(M_i)$ for $i=1,\dots,N$
\STATE $y_i \gets z_i + \sum_{j\neq i}\alpha_{ij} z_j + w_i$
\STATE $\hat{M}_i \gets D_{\phi_i}(y_i)$
\STATE $C_i \gets -\mathcal{L}[\hat{M}_i,M_i]$
\STATE $\phi_i \gets \phi_i - \eta\, \nabla_{\phi_i} C_i$ for $i=1,\dots,N$
\STATE $\theta_i \gets \theta_i - \eta\, \sum_{j=1}^{N} \nabla_{\theta_i} C_j$ for $i=1,\dots,N$
\ENDWHILE
\end{algorithmic}
\end{algorithm}

\begin{table}[t]
\centering
\caption{Per-receiver decoding complexity.}
\label{tab:decoder_complexity_bigO}
\resizebox{\columnwidth}{!}{%
\begin{tabular}{lcc}
\toprule
\textbf{Method} & \textbf{Decoder Type} & \textbf{Per-receiver Complexity} \\
\midrule
SiameseNet (Proposed) & Per-user MLP & $O(n)$ \\
Joint MLD & Exhaustive joint search & $O\left(n\,2^{kN}\right)$ \\
NOMA & SIC over user streams & $O\left(N\,C_{\mathrm{dec}}(n)+Nn\right)$ \\
RSMA & Common-stream SIC + private decoding & $O\left((N+1)\,C_{\mathrm{dec}}(n)+(N+1)n\right)$ \\
TDMA / OMA & Single-user decoding per slot & $O\left(C_{\mathrm{dec}}(n)\right)$ \\
\bottomrule
\end{tabular}%
}
\end{table}
\subsection{Decoder Complexity}
The decoder for user $i$ operates exclusively on $y_i\in\mathbb{R}^{n}$ with a single hidden layer of size proportional to $Nn$ ($n$ is the length of the codeword), yielding forward-pass complexity $\mathcal{O}(Nn)$, which scales linearly with $N$. In contrast, joint maximum-likelihood detection over all users requires $\mathcal{O}(2^{kN})$ evaluations (e.g., $2^{64}$ for $N=4$, $k=16$), and NOMA-SIC requires $N$ successive stages with inter-stage error propagation. Table~\ref{tab:decoder_complexity_bigO} summarises representative operation counts for $N\in\{2,4,8\}$.



\section{BLER Performance}
\label{sec: BLER Performance}
In this section, we carry out experiments on the systems mentioned in the aforementioned section. We study the
BLER performances of our models under different scenarios.
Our experiments use the following combinations $(k,n) : (4,8), (16,64), (256,2048)$, for $N=2, 4 \text{ and } 8$ respectively. These settings place the study in the finite-blocklength regime and span both short and medium blocklengths, allowing us to evaluate performance across a broader operating range rather than only the ultra-short regime. 

We vary, $E_b/N_0$ from 0 to 8~dB, and allow each transmitter to interfere with all other receivers given by~\eqref{eq: rx_eqn}. For fair spectral efficiency, we compare with a TDMA baseline where $N$ users transmit uncoded BPSK symbols per block at a coding rate of $ 1/N$ (Both systems have an overall rate of 1bps/Hz).

\subsection{Benchmarks and Baselines}
We compare the proposed SiameseNet design against two SIC-based reference schemes: power-domain non-orthogonal multiple access (NOMA) and rate-splitting multiple access (RSMA). In both baselines, the total transmit power is normalised as
\begin{equation}
\sum_{u=1}^{N} p_u = N,
\end{equation}
where $u$ denotes the user and $p_u$ denotes the transmit power of the $u^{th}$ user. We use the shorthand 
\begin{equation}
h_{i,j} =
\begin{cases}
1, & j=i,\\
\alpha, & j\neq i,
\end{cases}
\end{equation} 
so that the desired signal arrives with unit gain, while interfering users are scaled by the interference factor $\alpha$.

For fairness, each user transmits $k$ information bits and the baselines are configured to match the effective rate used by the SiameseNet design, $R = 1/N$.

\paragraph{NOMA baseline:}
For the NOMA benchmark, the received signal at receiver $i$ is
\begin{equation}
y_i=\sum_{j=1}^{N} h_{i,j}\sqrt{p_j}\,x_j+w_i,
\qquad w_i\sim\mathcal{N}(0,\sigma^2),
\end{equation}
where $x_j$ denotes the coded BPSK symbol stream of user $j$. Under perfectly symmetric received powers (Experiment I later in this section), a deterministic SIC order is not naturally defined. Therefore, for the NOMA baseline, we use a random SIC order over the interfering users.
We note that this choice removes one of \mbox{NOMA}'s
principal advantages: an optimised decoding order derived from
channel or power asymmetry~\cite{NOMA_Decoding}. This also implies that the optimal deployment of NOMA is contingent on a significant power differential between the 2 users, but our system is not constrained by such limitations.
Under symmetric power conditions, no such ordering is
naturally available, and the random order may underestimate
\mbox{NOMA}'s true performance relative to settings where a
power differential exists and a greedy decoding order can be
applied.

\color{black}

Let $\pi_i(1),\pi_i(2),\ldots, \pi_i(N-1)$ denote a random ordering of the interfering users at receiver $i$. Receiver $i$ successively decodes these interference streams, subtracts their reconstructed contributions, and then decodes its own stream. After cancelling the first $m-1$ decoded interference streams, the residual signal is
\begin{equation}
y_i^{(m)} = y_i - \sum_{r=1}^{m-1} h_{i,\pi_i(r)}\sqrt{p_{\pi_i(r)}}\,\hat{x}_{\pi_i(r)}.
\end{equation}
After cancelling all $N-1$ interfering streams, the desired stream is decoded from
\begin{equation}
\tilde{y}_i = y_i - \sum_{r=1}^{N-1} h_{i,\pi_i(r)}\sqrt{p_{\pi_i(r)}}\,\hat{x}_{\pi_i(r)}.
\end{equation}
The Log Likelihood Ratio (LLR) for the current BPSK symbol is then calculated from the residual signal, followed by channel decoding.

\paragraph{RSMA baseline:}
For the RSMA benchmark, each user splits its message into a common part and a private part. The transmitted signal of user $u$ is written as
\begin{equation}
s_u=\sqrt{\beta p_u}\,x_{u,c}+\sqrt{(1-\beta)p_u}\,x_{u,p},
\qquad 0\leq \beta \leq 1,
\end{equation}
where $x_{u,c}$ and $x_{u,p}$ denote the coded BPSK common and private streams, respectively, and $\beta$ controls the power split. The received signal at receiver $i$ is
\begin{equation}
y_i=\sum_{j=1}^{N} h_{i,j}s_j+w_i.
\end{equation}

Each user still transmits a total of $k$ information bits, now split as $k=k_c+k_p$ between common and private parts, with the overall effective rate matched to the SiameseNet design. At receiver $i$, the common streams are decoded first using SIC. Under symmetric power conditions, the common-stream decoding order is also taken to be random. Let $\rho_i(1),\rho_i(2),\ldots,\rho_i(N)$ denote the decoding order of the common streams. After cancelling the decoded common streams, the residual signal is
\begin{equation}
\bar{y}_i
=
y_i
-
\sum_{r=1}^{N}
h_{i,\rho_i(r)}
\sqrt{\beta p_{\rho_i(r)}}\,\hat{x}_{\rho_i(r),c}.
\end{equation}
Receiver $i$ then decodes its own private stream from
\begin{equation}
\bar{y}_i
=
\sqrt{(1-\beta)p_i}\,x_{i,p}
+
\alpha\sum_{j=1,\,j\neq i}^{N}\sqrt{(1-\beta)p_j}\,x_{j,p}
+
w_i.
\end{equation}
As with the NOMA baseline, the LLRs for the BPSK symbols are computed from the residual signal and passed to the channel decoder.
Thus, the NOMA baseline performs full SIC over the interfering users before decoding the desired stream, while the RSMA baseline first removes the common parts and then decodes the desired private stream in the presence of residual private-user interference.

NOMA and RSMA are more suited to cases with unequal interference strengths, thus,  for Experiment I, this should be treated as an indicative baseline. 

For Experiment I, we plot the Polyanskiy-Poor-Verdú (PPV) ~\cite{Poor_2010} normal approximation for an equivalent single-user AWGN channel as a finite-blocklength benchmark, which helps contextualise the performance gap introduced by multi-user interference.
We limit ourselves to only using PPV as an indicative bound, but a detailed quantitative analysis is not possible without a thorough information theoritic study of the system, but this is currently beyond the scope of the paper.
\color{black}

\subsection{Experiment I: Equal Interference from All Interferers}\label{subsec: Scenario I}
\begin{figure}[t]
\centering
\includegraphics[width=0.9\columnwidth]{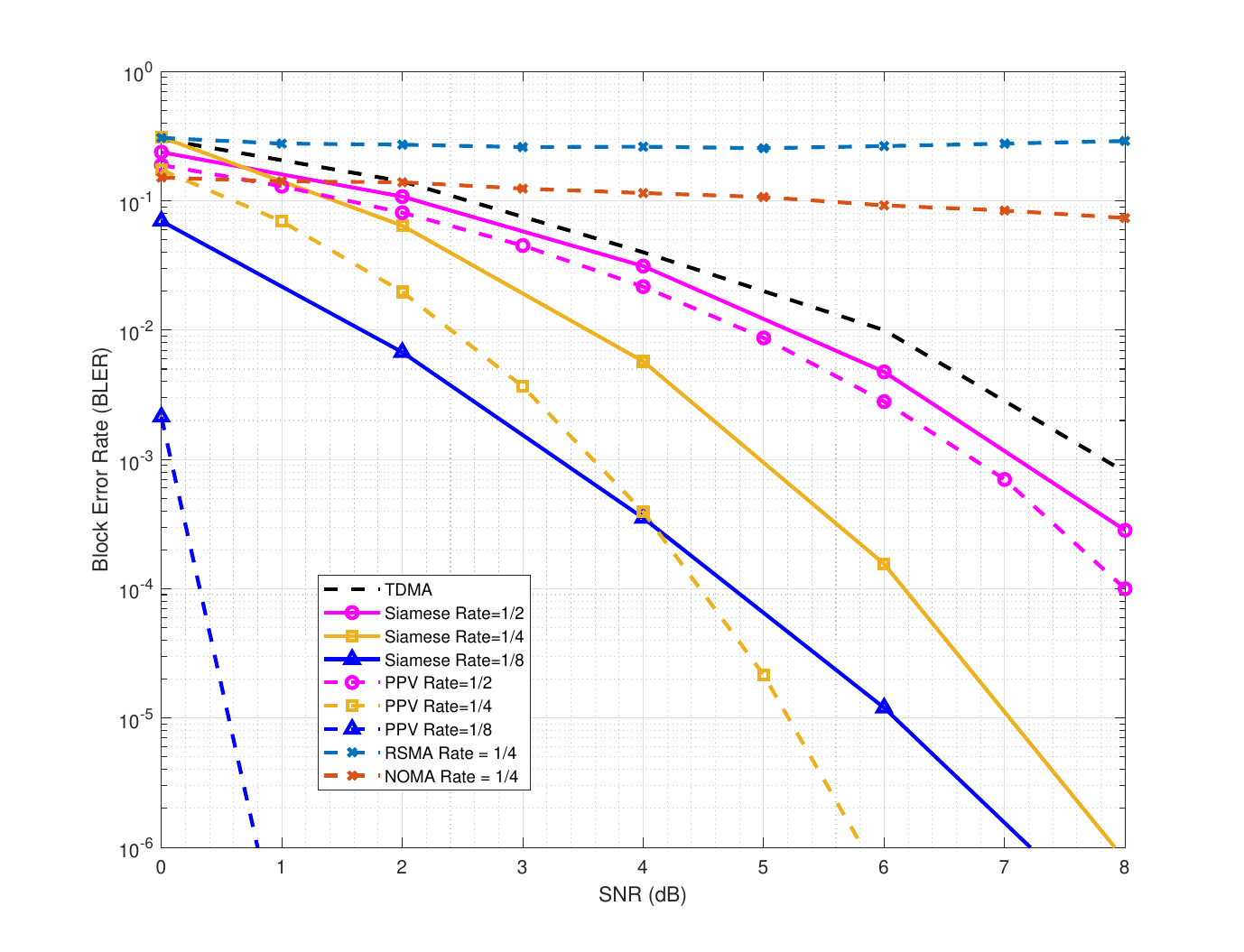}
\caption{BLER versus SNR for $N\in\{2,4,8\}$ under equal interference.}
\label{fig: BLER Performance: Varying user count}
\end{figure}

\begin{figure}[t]
\centering
\includegraphics[width=0.9\columnwidth]{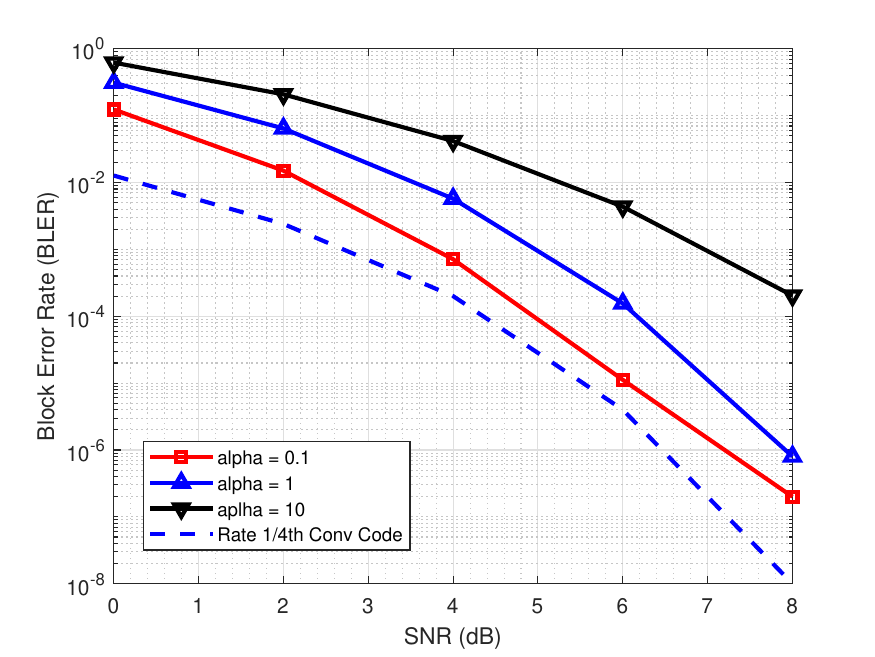}
\caption{Per-user BLER for the 4-user case.}
\label{fig: BLER Performance with varying alpha 4 users}
\end{figure}
\begin{figure}[t]
\centering
\includegraphics[width=0.9\columnwidth]{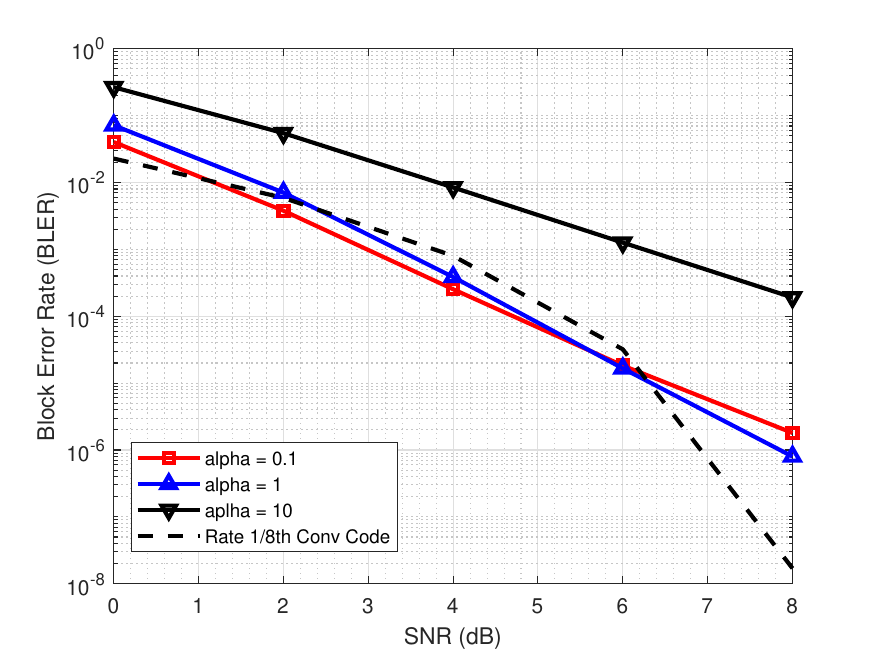}
\caption{Per-user BLER for the 8-user case.}
\label{fig: BLER Performance with varying alpha 8 users}
\end{figure}

\begin{figure}[t]
\centering
\includegraphics[width=0.9\linewidth]{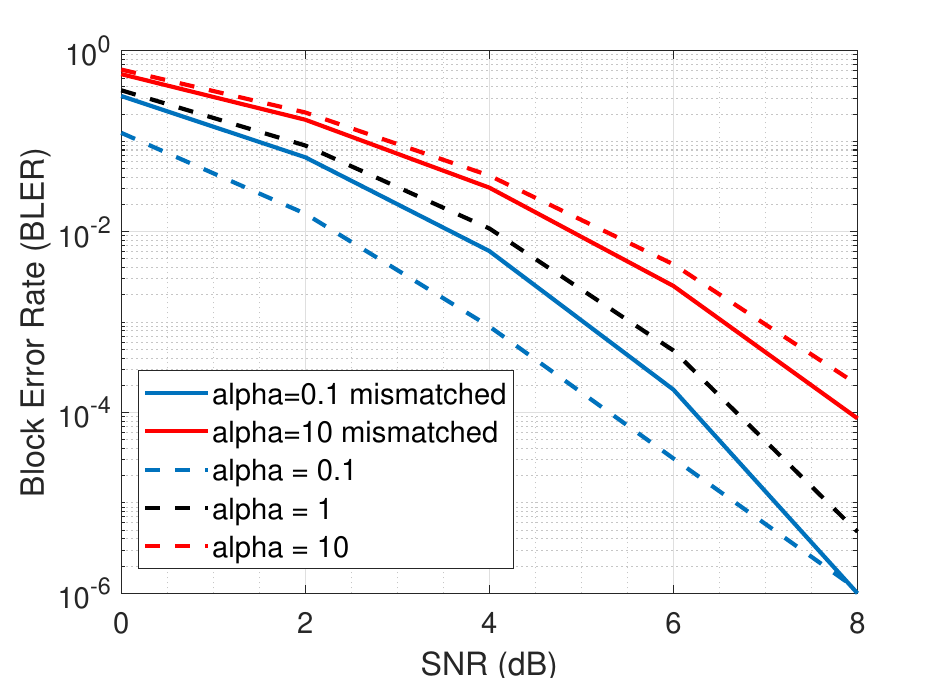}
\caption{BLER with mismatched $\alpha$ (trained at $\alpha{=}1$).}
\label{fig: BLER Performance with mismatched alpha}
\end{figure}

In this case, we assume all the signals are equally powerful and thus the interference strengths ($\alpha_{ij}$s) are equal for all users ($\alpha_{ij}=\alpha$ for all $i\neq j$). Fig.~\ref{fig: BLER Performance: Varying user count} shows that for $N\in\{2,4,8\}$ the Siamese network outperforms TDMA. A standard TDMA, SiameseNet 2 user with rate 1/2, 4 user with rate 1/4 and 8 user with rate 1/8 all have the same overall rate of 1 bps/Hz. Increasing $N$ lowers the per-user rate (from $1/2$ to $1/8$), implicitly increasing the minimum distance and improving BLER beyond a crossover SNR.
For 2 users we observe about $0.5$~dB gain; for 4 users the gain is about $2.5$~dB; the 8-user case improves further when compared with TDMA.  We discuss the mechanism of this behaviour in Section~\ref{sec: Latent Space Analysis}.

We also plot the corresponding PPV for Rates $1/2$ (2-user case), $1/4$ (4-user case) and $1/8$ (8-user case) and compare it with our results. 
Against the Rate=$1/2$ PPV, the performance is around $0.5$~dB worse for 2-User, $1.5$~dB worse for 4-User (Rate=$1/4$ PPV) and much worse for the 8-User.
This highlights that our system is more suited to short blocklength systems than larger blocklength systems.

\subsection{Experiment II: Mismatched $\alpha$: A Test for Robustness }\label{subsec: Scenario 2}
In this scenario, we explore the impact of a mismatch in $\alpha$.
We use the model trained for $\alpha =1$ and test it with different values of $\alpha$. 
Fig. \ref{fig: BLER Performance with mismatched alpha} shows the BLER performance with mismatched values of interferences. 
The BLER performance of the model (denoted by the solid blue line) is worse than what would be obtained with the correct model for $\alpha = 0.1$ (dashed blue line) , but the gap narrows with increasing SNR.
On the other hand, for $\alpha =10$, the mismatched model performance (denoted by the solid red line) performs as good as the true model (dashed red line).
This implies that our model has the ability to accommodate moderate errors in the estimation of the interference strength.
A more detailed analysis of the performance is presented in Sec~\ref{sec: Latent Space Analysis}.

\subsection{Experiment III: Unequal Interference Strengths}\label{subsec: scenario 3}
\begin{figure}[t]
\centering
\includegraphics[width=0.9\linewidth]{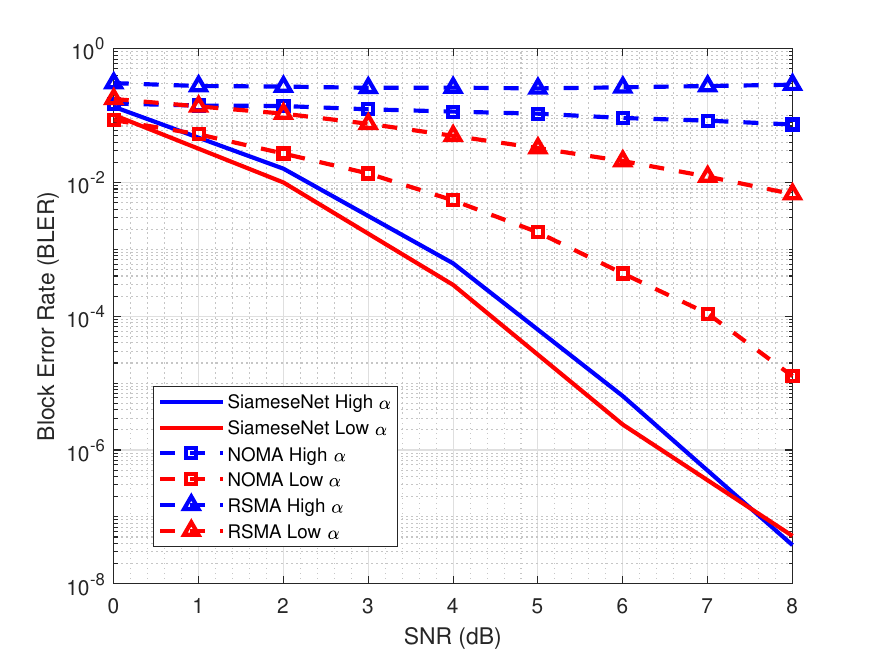}
\caption{BLER with unequal interference strengths.}
\label{fig: BLER Performance with unequal alpha}
\end{figure}

In this experiment, we examine the case when the interferers have different/unequal interference strengths.
We consider a scenario where $\alpha_{12} = \alpha_{21} = 1$, but  $\alpha_{13} = \alpha_{14} = \alpha_{23}=\alpha_{24}=\alpha_{34}=0.1$, that is, user 2 has a stronger interference with User 1 compared to users 3 and 4.
Fig.~\ref{fig: BLER Performance with unequal alpha} demonstrates the BLER performance of the users in this scenario.  
We Plot the high and low interferers for every scheme and compare it with our model. 
As we can observe, our model outperforms both NOMA and RSMA. A 
Further analysis of this ability needs a more thorough discussion, which is beyond the scope of this work and it will be explored in future.

\section{Preliminary results with MIMO}
\label{sec: Integration with MIMO}
\begin{figure}[t]
\centering
\includegraphics[width=0.9\columnwidth]{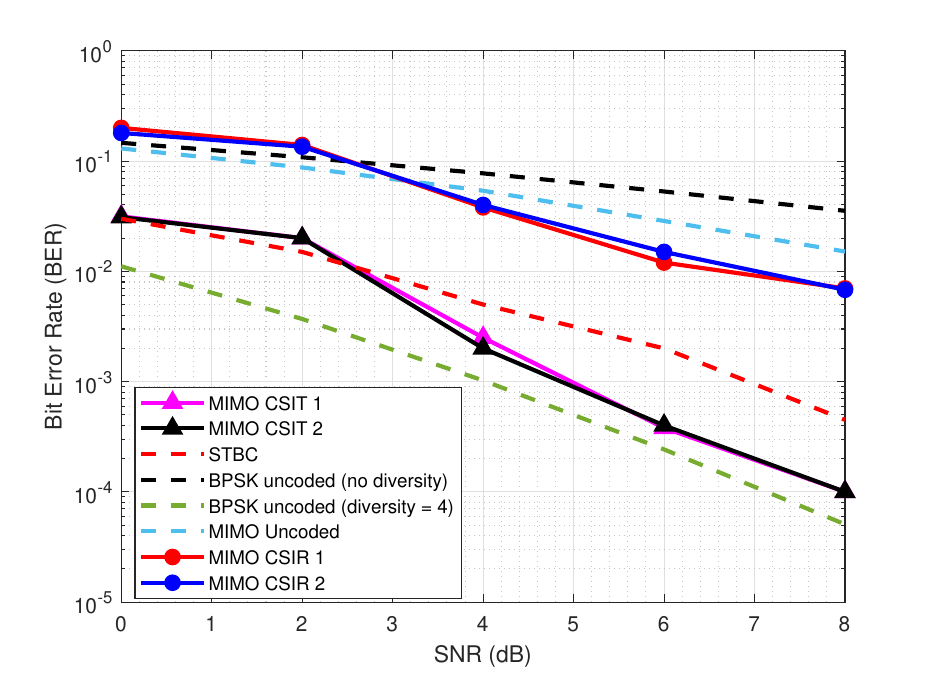}
\caption{BER for $2\times2$ MIMO under CSIT/CSIR compared to baselines.}
\label{fig: BER Performance MIMO}
\end{figure}

The results in this section are intended only as a proof-of-concept extension of the proposed framework to MIMO and do not constitute a general performance study over randomly varying channel ensembles. 
We study a $2\times2$ MIMO interference channel with i.i.d.\ complex Gaussian (Rayleigh) links and $\alpha{=}1$. Channels are drawn once and held fixed during training. The Models are trained for 50 epochs (batch size 64) over 50000 messages using categorical cross-entropy to reduce BER. For a 2-user MIMO scenario, the input is a 4-bit message at rate $1/2$, mapped to 8 symbols over two spatial streams (4 uses each), giving 1~bpcu per stream. Fig.~\ref{fig: BER Performance MIMO} reports BER under CSIT and CSIR.

\begin{figure}[t]
\centering
\includegraphics[width=0.9\columnwidth]{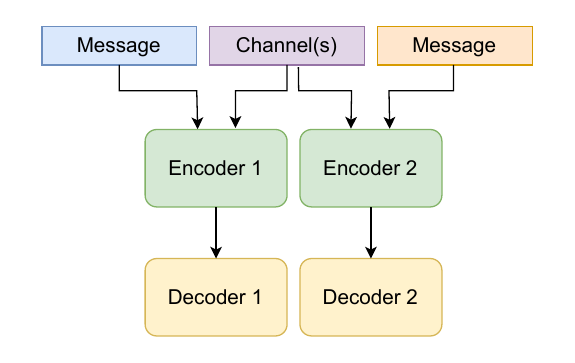}
\caption{Model variant for MIMO with CSIT (encoder knows the channel).}
\label{fig: MIMO_imple_CSIT}
\end{figure}

\textbf{Scenario I: CSIT.}
Channel matrices are available at the transmitters. We implement a model similar to~\cite{Paul_Intf_lim} with an additional encoder input for the channel (Fig.~\ref{fig: MIMO_imple_CSIT}), enabling beamforming and coded spatial multiplexing (akin to STBC~\cite{STBC}). We compare with single-user $2\times2$ Alamouti STBC~\cite{Alamouti} (fourth-order diversity) and single-user $2\times2$ MIMO with ML decoding (second-order). STBC has 1~bpcu; our learned code achieves 1~bpcu per user (8 bits over 2 TX and 8 uses) for two simultaneous users and exhibits diversity behaviour consistent with fourth order under the fixed-channel setup. Within this setting, CSIT trends outperform STBC and are within $\approx 0.5$~dB of uncoded BPSK with fourth-order diversity (Fig.~\ref{fig: BER Performance MIMO}).

\begin{figure}[t]
\centering
\includegraphics[width=0.9\columnwidth]{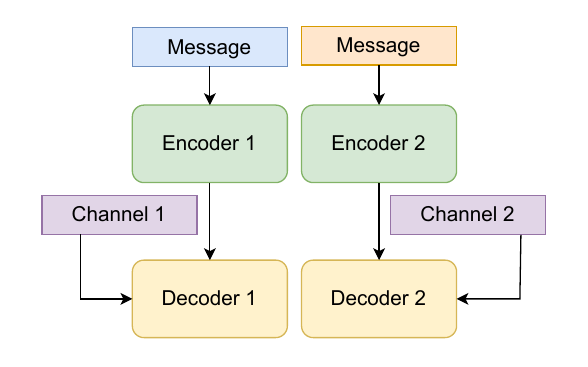}
\caption{Model variant for MIMO with CSIR (decoder knows the channel).}
\label{fig: MIMO_imple_CSIR}
\end{figure}

\textbf{Scenario II: CSIR.}
Channels are provided only to the decoders (Fig.~\ref{fig: MIMO_imple_CSIR}). The end-to-end model learns channel-agnostic codewords and suppresses cross-user interference via increased inter-user separation while retaining redundancy for coding. Under the fixed-channel experiments, diversity behaviour is consistent with second order, with about 2~dB improvement over uncoded $2\times2$ MIMO (Fig.~\ref{fig: BER Performance MIMO}).

\noindent\textit{Caveat (applies to all MIMO results in Section~V): Channels were drawn once and held fixed throughout training and evaluation. Therefore, the MIMO observations are preliminary and should not be interpreted as general performance over fading ensembles. In follow-up work, we will train and test over ensembles of randomly drawn Rayleigh realisations to evaluate channel-agnostic performance, diversity, and spectral efficiency.}
\color{black}
\begin{table}[H]
\centering
\caption{$d_{\min}$ for $\alpha\in\{0.1,1\}$ versus user count and effective rate.}
\label{tab:d-min table user count}
\begin{tabular}{|@{}l|c|c|c@{}|}
\hline
\textbf{User count, R}& \textbf{2-user, 1/2} & \textbf{4-user, 1/4} & \textbf{8-user, 1/8} \\
\hline
$\alpha = 1$ & 2.97 & 3.91 & 5.51 \\\hline
$\alpha = 0.1$ & 3.58 & 5.02 & 7.15 \\
\hline
\end{tabular}
\end{table}

\begin{table}[H]
    \centering
    \caption{Range of $\psi_{ii}$ for the $\alpha = 0.1$  and $\alpha=1$ cases.}    \begin{tabular}{|c|c|c|c|c|c|c|}\hline
         \textbf{User Count, R}&  \multicolumn{2}{|c|}{\textbf{2-user, 1/2}}& \multicolumn{2}{|c|}{\textbf{4-user, 1/4}}&\multicolumn{2}{|c|}{\textbf{8-user, 1/8}}\\\hline
 &  Mean&Range&  Mean&Range& Mean&Range\\\hline
         $\alpha =1$&  89.99&19-149& 89.99&18-152&89.99&10-140\\\hline
         $\alpha =0.1$&  89.99&22-146& 89.97&76-104&89.88&65-115\\ \hline
    \end{tabular}
    
    \label{tab:self angle table user count}
\end{table}
\begin{table}[H]
    \centering
    \caption{Range of $\psi_{ij}$ for the $\alpha = 0.1$  and $\alpha=1$ cases.}
    \begin{tabular}{|c|c|c|c|c|c|c|}\hline
         \textbf{User Count, R}&  \multicolumn{2}{|c|}{\textbf{2-User, 1/2}}& \multicolumn{2}{|c|}{\textbf{4-User, 1/4}}&\multicolumn{2}{|c|}{\textbf{8-user, 1/8}}\\\hline
 &  Mean&Range&  Mean&Range& Mean&Range\\\hline
         $\alpha =1$&  89.99&89-91& 90.00&88-92&89.99&88-92\\\hline
         $\alpha =0.1$&  89.99&37-142& 89.98&60-118&90.00&70-115\\ \hline
    \end{tabular}
    
    \label{tab:cross angle table user count}
\end{table}

\begin{figure}[t]
   \centering
    \includegraphics[width=0.9\linewidth]{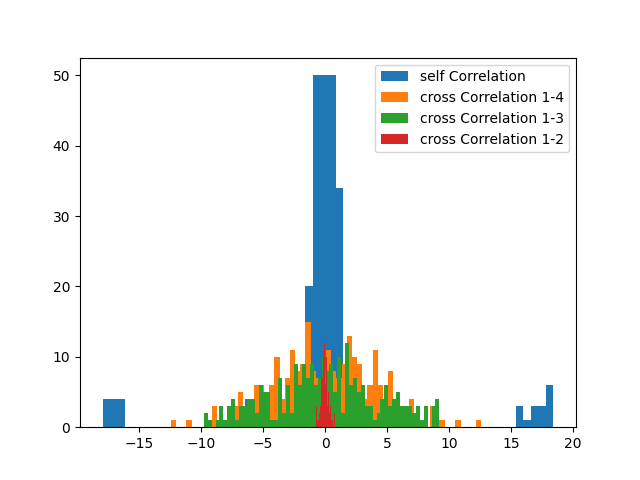}
    \caption{Correlation for 4 users $\alpha=1$ SiameseNet.}
    \label{fig: Corr unequal}
\end{figure}
\begin{figure}[t]
   \centering
    \includegraphics[width=0.9\linewidth]{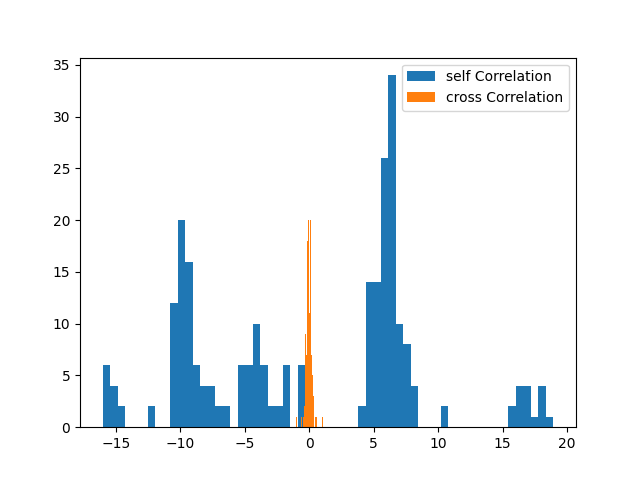}
    \caption{Correlation for 4 users $\alpha=1$ SiameseNet.}
    \label{fig: Corr 1 equal}
\end{figure}
\section{Latent Space Analysis}
\label{sec: Latent Space Analysis}
We analyse the encoders’ latent codewords to explain BLER via pairwise distances and correlations. The observed BLER trends are consistent with the learned inter-codeword separation, with the minimum distance  $d_{\min}$ emerging as the most informative geometric indicator in our experiments. Table~\ref{tab:d-min table user count} shows that $d_{\min}$ increases as the effective per-user rate decreases (e.g., from 2 to 8 users), with a stronger effect at $\alpha{=}0.1$, indicating that coding gain outweighs added interference.

Angles cluster near $90^\circ$ on average but depend on $\alpha$: same-user angles $\psi_{ii}$ are closer to $90^\circ$ for $\alpha{=}0.1$, while cross-user angles $\psi_{ij}$ approach $90^\circ$ more at $\alpha{=}1$ (Tables~\ref{tab:self angle table user count}, \ref{tab:cross angle table user count}).

Consistent with Section~\ref{sec: BLER Performance}, performance remains robust as interference scales because the model reduces cross-user correlation (Fig.~\ref{fig: Corr 1 equal}). Orthogonalisation is \emph{selective}: in Section~\ref{subsec: scenario 3} (Fig.~\ref{fig: Corr unequal}), user~1 approaches orthogonality relative to the dominant interferer (user~2; $\alpha{=}1$) but not to weaker interferers (users 3–4), instead keeping redundancy for coding. 

We do not design the model to explicitly reduce the correlation neither do we claim any guarantees on such a behaviour.
More detailed research is necessary to make concrete claims on this behaviour. But that is beyond the scope of the current paper.

\section{Conclusions}
\label{sec: conclusions}
This paper demonstrated that end-to-end deep learning can effectively address multi-user interference in finite-blocklength IoT networks.
By extending the two-user SiameseNet framework to
$4$ and $8$ users, we showed that the approach scales gracefully while
retaining the per-user decoder complexity, which scales logarithmically, whereas for joint detection, it scales exponentially. 
 
The framework also exhibits strong practical robustness: the performance degrades gracefully under interference-strength mismatch and outperforms both NOMA and RSMA under unequal interference strengths, scenarios directly relevant to heterogeneous IoT deployments.
Latent-space analysis provides a principled geometric insight into these
gains, with the minimum distance increasing as the effective per-user rate decreases, confirming that coding gain outweighs added interference as the number of users grows.

Preliminary $2\times 2$ MIMO results further suggest that the framework
extends naturally towards multi-antenna IoT gateways, with CSIT results
outperforming Alamouti STBC and CSIR results achieving second-order
diversity behaviour.
 
These results establish a strong foundation for end-to-end learned
transceivers in dense, interference-limited IoT scenarios.
Future work will pursue fading-ensemble MIMO evaluation, adaptive
interference-strength estimation, stronger IoT-oriented baselines, and an
information-theoretic analysis of the learned code geometry.
\color{black}
\section{Acknowledgments}
The authors of this paper would like to thank Prof Nambi Seshadri of UC San Diego and Prof R David Koilpillai of IIT Madras for their guidance and support.

\bibliographystyle{IEEEtran}
\bibliography{Bibliography}

\end{document}